\documentclass{PoS}

\usepackage{wrapfigure}

\title{A new detector for deep inelastic physics}

\ShortTitle{A new detector for deep inelastic physics}

\author{Peter Kostka\\
  University of Liverpool, Liverpool, L69 7ZE, Great Britain\\
  Email: \email{peter.kostka@liv.ac.uk}}

\author{Alessandro Polini\\
  INFN Bologna, via Irnerio 46, 40126 Bologna, Italy\\
  E-mail: \email{alessandro.polini@bo.infn.it}}

\author{\speaker{David M. South}\thanks{on behalf of the LHeC Collaboration.}\\
  Deutsches Elektronen Synchrotron, Notkestrasse 85, 22607 Hamburg, Germany\\
  E-mail: \email{david.south@desy.de}}

\abstract{The Large Hadron Electron Collider (LHeC) is a proposed upgrade to the LHC,
  to provide high energy, high luminosity electron-proton and electron-ion collisions to run
  concurrently with Phase $2$ of the LHC. The key elements of the LHeC detector
  and the requirements from the physics programme are outlined, followed by a
  brief description of the baseline LHeC detector design.}

\FullConference{The European Physical Society Conference on High Energy Physics -EPS-HEP2013\\
                18-24 July 2013\\
                Stockholm, Sweden}

\begin{document}

\section{Introduction to the LHeC}

The LHeC~\cite{lheccdr} is a planned $ep$ collider at CERN, where one of the $7$~TeV LHC proton
beams is brought into collision with a $60$~GeV electron beam, operating with a design luminosity
of about $10^{33}$~cm$^{-2}$~s$^{-1}$.
Data taking at the LHeC is to run simultaneously to the main LHC $pp$ collisions during Phase
$2$ operation (beyond 2023).
The preferred accelerator layout is a linac-ring design, employing two $1$km long linacs,
with energy recovery for the lepton beam; an alternative ring-ring approach has also been
considered~\cite{oliverDIS}.
The LHeC combines high energy, high precision and high luminosity in a wide reaching physics programme
to complement that of the LHC, aspects of which are detailed in other contributions to these proceedings
or elsewhere~\cite{maxDIS}.
The requirements of a proposed LHeC detector, driven by the goals of the physics programme, are outlined
below, followed by a brief description of main components of the baseline detector design.

\section{Detector requirements}
\label{sec:design}

The LHeC detector needs to be designed, constructed and ready for use at the beginning of LHC Phase $2$,
in approximately $12$ years from now, thus prohibiting a dedicated, large scale R\&D programme.
Such a detector, which should be modular and flexible in design and be achievable at a reasonable cost,
can however benefit from current and upgrade LHC and ILC technologies, as well as from the experience
gained at HERA.
It must be able to run concurrently with the other LHC $pp$ and $pA$ experiments, in order to make
use of the LHC beam and record the corresponding $ep$ and $eA$ data, and would be located at Point~$2$,
the only interaction point of the LHeC beams.
%
%
%
To fulfil all facets of the experimental programme, the following physics requirements are desirable:

{\footnotesize
\begin{itemize}
\item A high resolution tracking system to provide excellent primary vertex resolution and resolution of secondary
vertices down to small angles in forward direction for high $x$ heavy flavour physics and searches.
\item A precise $P_{T}$ measurement, matched to calorimeter signals calibrated and aligned to an accuracy of  $1$~mrad.
\item Full coverage calorimetry with an electron energy measured to $10\% / \sqrt{E}$, calibrated using the kinematic peak
and double angle method to the per-mil level and hadronic energy measured to $40\% / \sqrt{E}$, with a calibrated
$P_{T}$ balance to an accuracy of $1$\%.
\item A muon system, for tagging and momentum measurement in combination with tracking
\item Tagging of backward scattered photons and electrons for a precise measurement of luminosity and photoproduction physics.
\item Tagging of forward scattered protons, neutrons and deuterons to fully investigate diffractive and deuteron physics.
\end{itemize}
}

The proposed baseline detector design to achieve these goals is briefly described below, more details can be found
in the LHeC Conceptual Design Report~\cite{lheccdr}.

\section{The baseline LHeC detector design}
\label{sec:detector}

\begin{wrapfigure}{r}{95mm}
  \begin{center}
    \includegraphics[width=0.6\textwidth]{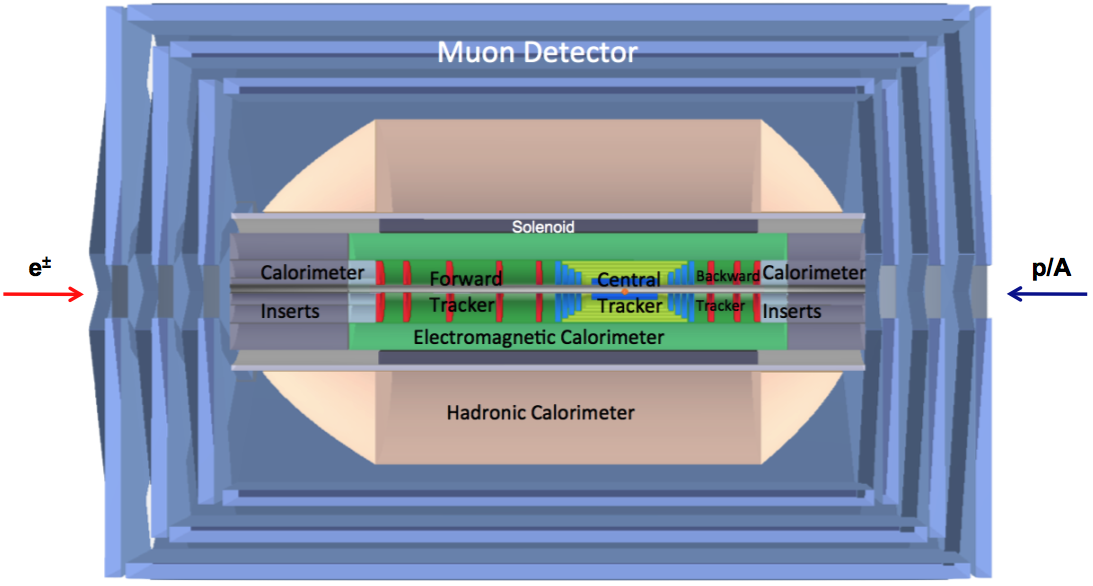}
  \end{center}
  \vspace{-0.4cm}
  \caption{An $r$-$z$ cross section of the main LHeC detector in the baseline configuration with the solenoid
    and dipoles placed between the EMC and the HAC. The proton beam, from the right, collides with the
    electron beam, from the left, at the IP which is surrounded by a central tracking system complemented
    by larger forward and backward tracking detectors and followed by sets of calorimeters and a muon detector.
    Additional systems, not pictured, are located either side of the main detector in the LHeC tunnel.}
 \label{fig:centraldet}
\end{wrapfigure}

The LHeC detector provides hermetic coverage, up to the very forward and backward directions to provide
a precise energy measurement and an intrinsic asymmetry, reflecting the corresponding asymmetry in the
beam energies.
Figure~\ref{fig:centraldet} shows the baseline design: the dimensions of the main LHeC detector are
$14$m $\times 9$m, which is much smaller than the CMS ($21$m $\times 15$m) or
ATLAS ($45$m $\times 25$m) detectors.


The LHeC tracking detector has a high acceptance, a compact tracking design and is completely contained~within
the electromagnetic calorimeter (EMC).
An all silicon design using pixel and strip detectors is employed, with more coverage in the proton direction
and an elliptical arrangement of those layers closest to the similarly shaped beampipe.
Transverse momentum resolutions down to $10^{-3}~{\rm GeV}^{-1}$ and impact parameter resolutions of
distances as small as $10~\mu m$ are expected from simulation~\cite{lheccdr}.


The main EMC in the barrel region, $2.8 < \eta< -2.3$, is based on the LAr/Pb design used in ATLAS
and employs three different granularity sections longitudinally, equivalent to $25-30$ radiation lengths.
The baseline hadronic calorimeter (HAC) design uses steel absorber and scintillator sampling plates,
similar to the TILE calorimeter in ATLAS, with a depth of $7-9$ interaction lengths.
Complementing the main calorimeters are electromagnetic and hadronic inserts in both the forward
and backward regions.
The baseline magnet design is a small $3.5$T coil between the EMC and HAC, placing the solenoid along
with dipoles needed for the steering of the electron beam required for head on collisions, conveniently
within the same cold vacuum vessel.


A muon system consisting of $2$-$3$ layers is included, each with a double trigger layer and a layer for 
measurements, comprised of thin gap chambers, resistive plate chambers and drift tubes.
Muon momenta is measured by the inner tracker, in combination with signals from the muon system.
The luminosity is measured using Bethe-Heitler collinear photons detected at $z \approx -120$~m,
as well as the analysis of QED Compton events.
Particle taggers are foreseen both in the backward ($z \approx -6$~m and $-62$~m) and forward
($z \approx +100$~m and $+420$~m) regions.
%

\end{document}